\documentclass[prb,aps,showpacs,twocolumn]{revtex4}

\usepackage{amsmath}
\usepackage{bm}
\usepackage{graphicx}

\begin{document}

\title{Order--disorder oscillations in exciton--polariton superfluids}

\author{Hiroki Saito}
\affiliation{Department of Engineering Science, University of
Electro-Communications, Tokyo 182-8585, Japan}

\author{Tomohiko Aioi}
\affiliation{Department of Engineering Science, University of
Electro-Communications, Tokyo 182-8585, Japan}

\author{Tsuyoshi Kadokura}
\affiliation{Department of Engineering Science, University of
Electro-Communications, Tokyo 182-8585, Japan}

\date{\today}

\begin{abstract}
The dynamics of an exciton--polariton superfluid resonantly pumped in a
semiconductor microcavity are investigated by mean-field theory.
Modulational instability develops into crystalline order and then ordered
and disordered states alternately form.
A supersolid-like state is also found, in which superflow coexists with
crystalline order at rest.
\end{abstract}


\maketitle

\section{Introduction}

Transitions between ordered and disordered states can be controlled by
changing the parameters of a system.
For example, a liquid can be solidified and then melted back into a liquid
by controlling the temperature or pressure.
Another example is a flow pattern in the wake of an obstacle.
A periodic vortex street can change into turbulence by increasing the
Reynolds number and turbulence can transform back into a vortex street by
decreasing the Reynolds number.
In a large complex system, order and disorder can alternately appear in a
partial system, such as clouds in the sky.
However, in a simple laboratory system, such repeated transitions between
ordered and disordered states usually do not occur unless an external
parameter is changed~\cite{Collet}.
Ordering phenomena or pattern formations occur in a wide variety of
nonequilibrium dissipative systems~\cite{Cross}, such as B\'enard
convection cells~\cite{Benard} and chemical patterns in reaction
diffusion systems~\cite{Zaikin,Ouyang}.
However, once a pattern has been generated in these systems, it is
preserved if the parameters are fixed, although it may vary with time.
By contrast, we here propose a system in which crystalline order and
disorder alternately emerge despite external parameters being fixed.

The system that we consider is an exciton--polariton superfluid in a
planar semiconductor microcavity~\cite{Deng10,Carusotto12}.
There have been many recent experimental studies of this system,
including:
bistable properties~\cite{Baas,Gippius},
Bose--Einstein condensation~\cite{Kasprzak,Balili}, 
Bogoliubov excitations~\cite{Utsunomiya},
quantized vortices~\cite{Lagoudakis,Sanvitto},
superfluidity~\cite{Amo09},
quantum hydrodynamics~\cite{Nardin,Amo11},
and pattern formation~\cite{Sanvitto06,Manni}.
Of the various ways to create a polariton superfluid, we consider a system
of polaritons pumped by a quasi-resonant continuous-wave laser.
Since polaritons have a short lifetime of $\sim 10$ ps, which is
comparable to the time scale of the dynamics, the system is a
nonequilibrium dissipative system with continuous feeding.
It is thus suitable for studying nonlinear phenomena far from
equilibrium.

In this paper, we investigate pattern formation and destruction dynamics
in an exciton--polariton superfluid.
The polaritons are first pumped to the lower branch of the
bistability~\cite{Baas,Gippius}, which is an unstable stationary
state~\cite{Carusotto}.
Modulational instability then arises and a hexagonal crystalline pattern
develops, which eventually collapses, destroying the order.
However, the pattern reappears from the disordered distribution and grows
again.
Subsequently, the ordered and disordered distributions emerge
alternately.
We also show that the crystalline order can remain at rest even in the
presence of superflow.

This paper is organized as follows.
Section~\ref{s:form} formulates the problem,
Section~\ref{s:result} demonstrates numerical results,
and Section~\ref{s:conc} gives conclusions of this study.

\section{Formulation of the problem}
\label{s:form}

Excitons in quantum wells and photons confined between Bragg mirrors are
assumed to be described by macroscopic wave functions $\psi_{\rm X}$ and
$\psi_{\rm C}$, obeying the two-dimensional nonlinear Schr\"odinger
equations given by~\cite{Carusotto}
\begin{subequations} \label{GP}
\begin{eqnarray}
i \hbar \frac{\partial\psi_{\rm X}}{\partial t} & = & \hat H_0^{\rm X}
\psi_{\rm X} + \hbar \Omega_{\rm R} \psi_{\rm C} + g |\psi_{\rm X}|^2
\psi_{\rm X} - i \hbar \frac{\gamma_{\rm X}}{2} \psi_{\rm X}, \nonumber \\
\label{GPc} \\
i \hbar \frac{\partial\psi_{\rm C}}{\partial t} & = & \hat H_0^{\rm C}
\psi_{\rm C} + \hbar \Omega_{\rm R} \psi_{\rm X}
- i \hbar \frac{\gamma_{\rm C}}{2} \psi_{\rm C} \nonumber \\
& & + \hbar F(\bm{r})
e^{i (\bm{k}_{\rm p} \cdot \bm{r} - \omega_{\rm p} t)},
\nonumber \\
\end{eqnarray}
\end{subequations}
where $\hat H_0^j$ and $\gamma_j$ are respectively the free Hamiltonians
and the decay rates of an exciton ($j = {\rm X}$) and a photon ($j = {\rm
C}$), $\Omega_{\rm R}$ is the Rabi frequency, and $g$ is the
exciton--exciton interaction coefficient.
The free Hamiltonians are approximated as $\hat H_0^{\rm X} = \hbar
\omega_0^{\rm X}$ and $\hat H_0^{\rm C} = \hbar \omega_0^{\rm C} - \hbar^2
\nabla^2 / (2 m_{\rm C})$, where $m_{\rm C}$ is the in-plane effective
mass of a cavity photon.
Diagonalizing the first and second terms on the right-hand side (rhs) of
Eq.~(\ref{GP}), we obtain the dispersion relations $\omega_\pm(k)$ for the
upper and lower free polaritons.
The last term on the rhs of Eq.~(\ref{GPc}) describes pumping by an
external laser beam with a profile $F(\bm{r})$, in-plane wave vector
$\bm{k}_{\rm p}$, and frequency $\omega_{\rm p}$.

We numerically solve Eq.~(\ref{GP}) using the pseudospectral
method~\cite{Recipes}.
As an initial state, small white noise is set to $\psi_{\rm X}$ and
$\psi_{\rm C}$ to exclude exact numerical symmetry.
In the following calculations, we only consider the case of normal
incidence of the pump beam on the sample, i.e., $\bm{k}_{\rm p} = 0$.
The detuning of the pump frequency from the lower polariton is defined as
$\delta = \omega_{\rm p} - \omega_-(0)$.
We use the Gaussian profile of the pump beam as
\begin{equation} \label{gaussian}
F(\bm{r}) = F_0 \exp\left[ -(x^2 + y^2) / d^2 \right],
\end{equation}
where $F_0$ is proportional to the peak intensity and $d$ is the $1/e$
width.
The parameters are $m_{\rm C} = 2 \times 10^{-5} m_{\rm e}$ with $m_{\rm
e}$ being the electron mass, $\hbar \Omega_{\rm R} = 5$ meV, $g = 0.01
{\rm meV} \mu {\rm m}^2$, and $\hbar \delta = 1.53$ meV.
For simplicity, we assume $\omega_0^{\rm X} = \omega_0^{C}$ and
$\gamma \equiv \gamma_{\rm X} = \gamma_{\rm C} = (10 {\rm ps})^{-1}$.

\section{Numerical results}
\label{s:result}

\begin{figure}[tbp]
\includegraphics[width=8cm]{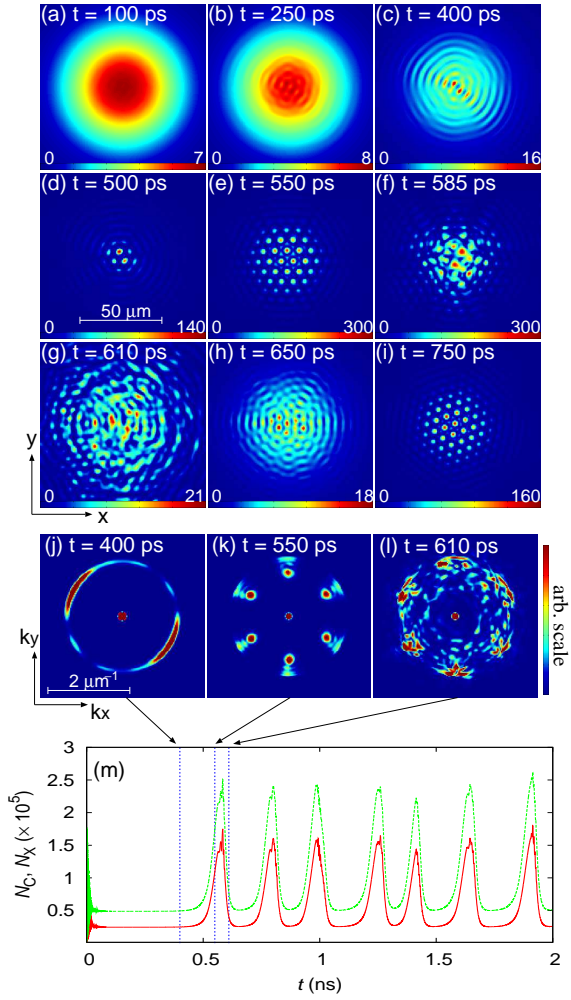}
\caption{
(color online) (a)--(i) Snapshots of the density profile $|\psi_{\rm
C}|^2$ of photons.
The pump parameters in Eq.~(\ref{gaussian}) are $\hbar F_0 = 8.97$ meV and
$d = 50$ $\mu{\rm m}$.
The color scale in each image is normalized by the peak density, where the
unit of density is $\mu{\rm m}^{-2}$.
The field of view of each panel is $100 \times 100$ $\mu{\rm m}$.
(j)--(l) Fourier-transformed profiles $|\int \psi_{\rm
C} e^{-i \bm{k} \cdot \bm{r}} d\bm{r}|^2$, where the field of view is $4
\times 4$ $\mu{\rm m}^{-1}$.
(m) Time evolution of $N_{\rm C} = \int |\psi_{\rm C}|^2 d\bm{r}$ (red
solid line) and $N_{\rm X} = \int |\psi_{\rm X}|^2 d\bm{r}$ (green dashed
line).
The vertical dotted lines show $t = 400$, $550$, and $610$ ps.
See supplemental material~\cite{movie}.
}
\label{f:main}
\end{figure}
Figures~\ref{f:main}(a)--\ref{f:main}(i) show the dynamics of the density
profile $|\psi_{\rm C}(\bm{r})|^2$ of photons ($|\psi_{\rm X}(\bm{r})|^2$
is always similar to $|\psi_{\rm C}(\bm{r})|^2$; it will not be shown
henceforth).
First, the polariton density grows in the region of $r \lesssim d$ and
reaches a steady state with a Gaussian profile (Fig.~\ref{f:main}(a)),
reflecting the pump profile in Eq.~(\ref{gaussian}).
Modulational instability then arises around the center of the pump region
and a pattern develops (Figs.~\ref{f:main}(b) and \ref{f:main}(c)).
At this stage, the pattern has no crystalline symmetry.
At $t \simeq 500$ ps, crystalline order emerges around the center and it
grows outward to form a triangular lattice (Figs.~\ref{f:main}(d) and
\ref{f:main}(e)).
The wave function has almost the same phase at each density peak.
This ordered state is not a steady state and the density of each peak
continues to increase.
When the peak density reaches some critical value, the triangular lattice
abruptly collapses, destroying the crystalline order
(Figs.~\ref{f:main}(f) and \ref{f:main}(g)).
The disturbance decays with time and a similar modulational pattern to
that in Fig.~\ref{f:main}(c) re-emerges (Fig.~\ref{f:main}(h)), which is
followed by the reappearance of crystalline order (Fig.~\ref{f:main}(i)).
The collapse and restoration of crystalline order are subsequently
repeated.

Figure~\ref{f:main}(m) shows the time evolution of the norms $N_{\rm C} =
\int |\psi_{\rm C}|^2 d\bm{r}$ and $N_{\rm X} = \int |\psi_{\rm X}|^2
d\bm{r}$.
After the initial fluctuations for $t \lesssim 100$ ps, $N_{\rm C}$ and
$N_{\rm X}$ reach plateaus corresponding to the steady state in
Fig.~\ref{f:main}(a).
The Fourier-transformed profile of this steady state localizes within $|k|
\lesssim 0.1$ $\mu{\rm m}^{-1}$ (data not shown).
As the modulational instability grows, a ring-shaped profile emerges at
$|k| \simeq 1.3$ $\mu{\rm m}^{-1}$ $\simeq 2 \pi / (5 \mu{\rm m})$
(Fig.~\ref{f:main}(j)), which reflects the pattern in
Fig.~\ref{f:main}(c).
The anisotropic shape of the ring depends on the small noise added to the
initial state.
As the crystalline order grows, $N_{\rm C}$ and $N_{\rm X}$ increase
rapidly and the ring in Fourier space changes to hexagonal peaks
(Fig.~\ref{f:main}(k)).
Since the peaks in Fig.~\ref{f:main}(k) are closer to the origin than the
ring in Fig.~\ref{f:main}(j), the period of the triangular lattice in
Fig.~\ref{f:main}(e) is larger than the modulation wavelength in
Fig.~\ref{f:main}(c).
The sudden reduction in $N_{\rm C}$ and $N_{\rm X}$ corresponds to the
collapse of the triangular lattice, where the six hexagonal peaks change
to a disordered distribution in Fourier space (Fig.~\ref{f:main}(l)).
The disordered components are subsequently cleaned up and the ring shape
grows, which is followed by hexagonal peaks.

To qualitatively understand the behaviors shown in Fig.~\ref{f:main}, we
consider the lower-polariton approximation.
Neglecting the upper polariton branch, we can reduce Eq.~(\ref{GP})
to~\cite{Carusotto}
\begin{eqnarray} \label{GPLP}
i \hbar \frac{\partial\psi_{\rm LP}}{\partial t} & = & \left(
-\frac{\hbar^2}{2 m_{\rm LP}} \nabla^2 + g_{\rm LP} |\psi_{\rm LP}|^2
- i \hbar \frac{\gamma}{2} \right) \psi_{\rm LP} \nonumber \\
& &  + \hbar F_{\rm LP}
e^{-i \omega_{\rm p} t},
\end{eqnarray}
where $\psi_{\rm LP} = X_{\rm LP} \psi_{\rm X} + C_{\rm LP} \psi_{\rm C}$,
$m_{\rm LP} = 2 m_{\rm C}$, $g_{\rm LP} = |X_{\rm LP}|^4 g$, and $F_{\rm
LP} = C_{\rm LP} F$ are the macroscopic wave function, the effective mass,
the interaction coefficient, and the pump intensity of the lower
polaritons, respectively.
$X_{\rm LP}$ and $C_{\rm LP}$ are the Hopfield coefficients.
For simplicity, we consider the case of homogeneous pumping and the
polariton wave function can be divided into
\begin{equation} \label{psi}
\psi_{\rm LP}(\bm{r}, t) = e^{-i \omega_{\rm p} t} [\phi_{\rm ss} +
\delta\phi(\bm{r}, t)].
\end{equation}
Substituting Eq.~(\ref{psi}) into Eq.~(\ref{GPLP}), we find that the
homogeneous steady state $\phi_{\rm ss}$ is determined by
\begin{equation} \label{ss}
\left( \hbar\delta - g_{\rm LP} |\phi_{\rm ss}|^2 + i \hbar \gamma / 2
\right) \phi_{\rm ss} = \hbar F_{\rm LP},
\end{equation}
which will have multiple solutions of $\phi_{\rm ss}$ for a certain range
of $F_{\rm LP}$ if $\delta / \gamma > \sqrt{3} / 2$.
For the parameters in Fig.~\ref{f:main}, $\delta / \gamma \simeq 23$ and
Eq.~(\ref{ss}) has three solutions for 0.8 meV $\lesssim \hbar F_{\rm LP}
\lesssim 14$ meV.
The steady state in Fig.~\ref{f:main}(a) corresponds to the state with
the lowest density of the three solutions.

We use Bogoliubov analysis to investigate the stability of the steady
state.
The first order of $\delta\phi$ in Eq.~(\ref{GPLP}) gives the
Bogoliubov--de Gennes equation.
For an excitation of the form $\delta\phi(\bm{r}, t) = u_k e^{-i\omega t
+ i \bm{k} \cdot \bm{r}} + v_k^* e^{i \omega^* t - i \bm{k} \cdot
\bm{r}}$, the Bogoliubov--de Gennes equation can be diagonalized to yield
the excitation frequency:
\begin{equation} \label{omega}
\omega = -i \frac{\gamma}{2} \pm \sqrt{ (\omega_k - \delta + \omega_{\rm
ss}) (\omega_k - \delta + 3 \omega_{\rm ss})},
\end{equation}
where $\omega_k = \hbar k^2 / (2m_{\rm LP})$ and $\omega_{\rm ss} = g_{\rm
LP} |\phi_{\rm ss}|^2 / \hbar$.
When the imaginary part of $\omega$ is positive, the corresponding mode
grows exponentially.
From Eq.~(\ref{omega}), the condition for this modulational instability
to occur is found to be
$\omega_{\rm ss} > \gamma / 2$ and $\delta - 2 \omega_{\rm ss} +
(\omega_{\rm ss}^2 - \gamma^2 / 4)^{1/2} > 0$,
which is satisfied when $r \lesssim 20$ $\mu{\rm m}$ in the situation
shown in Fig.~\ref{f:main}(a).
For $\omega_k = \delta - 2 \omega_{\rm ss}$, Eq.~(\ref{omega}) has the
largest imaginary part, which corresponds to the most unstable mode.
For the parameters in Fig.~\ref{f:main}(a), the most unstable wavelength
is estimated to be $\simeq 7$ $\mu{\rm m}$, which is in reasonable
agreement with the modulation in Fig.~\ref{f:main}(b).

The hexagonal crystalline order shown in Figs.~\ref{f:main}(e) and
\ref{f:main} (i) is a manifestation of the nonlinear effect.
To see this, we simplify the problem using a wave function with the form
\begin{equation} \label{model}
\psi_{\rm LP}(\bm{r}, t) = e^{-i \omega_{\rm p} t} \phi_{\rm ss} [1 + c(t)
f(\bm{r})],
\end{equation}
where $f(\bm{r}) = \sum_i \cos \bm{k}_i \cdot \bm{r}$.
Substituting Eq.~(\ref{model}) into Eq.~(\ref{GPLP}) and extracting terms
proportional to $f(\bm{r})$, we obtain
\begin{eqnarray} \label{ceq}
i \dot{c} & = & \left( \omega_k - \omega_{\rm p} - i \gamma / 2 \right) c
+ \omega_{\rm ss} \bigr[ 2 c + c^* + A (2 |c|^2 + c^2)
\nonumber \\
& & + B |c|^2 c \bigl],
\end{eqnarray}
where $A$ and $B$ are constants depending on $f(\bm{r})$.
The $O(c)$ terms in Eq.~(\ref{ceq}), which give Eq.~(\ref{omega}), are
responsible for the exponential growth of $c$ and we restrict the wave
number $|\bm{k}_i|$ to be the most unstable one.
If $A \neq 0$, $O(c^2)$ terms will become important as $c$ grows.
Equation~(\ref{ceq}) gives
\begin{equation}
\frac{d}{dt} |c|^2 = -\gamma |c|^2 + i \omega_{\rm ss} (c^2 - c^{*2})+ i
\omega_{\rm ss} A |c|^2 (c - c^*),
\end{equation}
which indicates that the solution satisfying $i A (c - c^*) > 0$ dominates
the exponential growth, breaking the symmetry with respect to $c
\rightarrow -c$ in the linear part of Eq.~(\ref{ceq}).
The simplest form of $f(\bm{r})$ with nonzero $A$ is $f(\bm{r}) =
\cos\bm{k}_1 \cdot \bm{r} + \cos\bm{k}_2 \cdot \bm{r} + \cos\bm{k}_3 \cdot
\bm{r}$ with $\bm{k}_1 + \bm{k}_2 + \bm{k}_3 = 0$; i.e., the three wave
vectors form a regular triangle.
In this case, $A = 1$ and $B = 15 / 4$.
Thus, the hexagonal order dominates the growth as the modulation amplitude
grows.
The term of $B$ in Eq.~(\ref{ceq}) effectively shifts the resonance by
$\omega_{\rm ss} B |c|^2$.
This is why the peak wave number decreases (Fig.~\ref{f:main}(k)) as the
hexagonal pattern grows.

\begin{figure}[tbp]
\includegraphics[width=8cm]{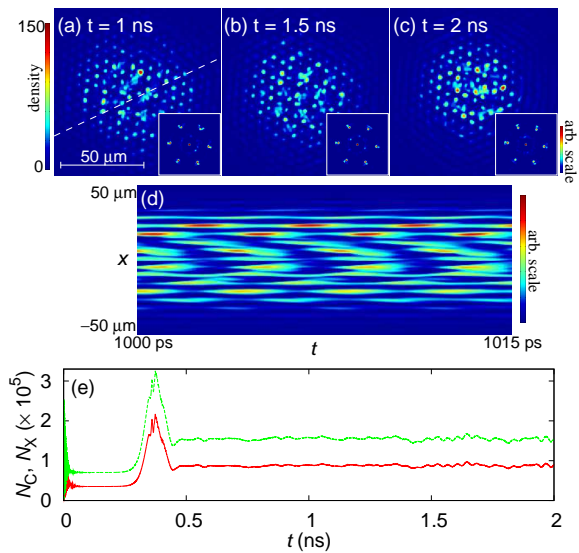}
\caption{
(color online) (a)--(c) Snapshots of the density profile $|\psi_{\rm C}|^2$
for $d = 60$ $\mu{\rm m}$.
The insets show Fourier-transformed profiles.
The field of view of the main panels is $100 \times 100$ $\mu{\rm m}$ and
that of the insets is $4 \times 4$ $\mu{\rm m}^{-1}$.
(d) Spatiotemporal image of the density on the dashed line in (a).
(e) Time evolution of the norms $N_{\rm C}$ (red solid line) and $N_{\rm
X}$ (green dashed line).
In (a)--(e), all the parameters except $d$ are the same as those in
Fig.~\ref{f:main}.
See supplemental material~\cite{movie}.
}
\label{f:local}
\end{figure}
We return to the numerical simulations of Eq.~(\ref{GP}).
When the peak density grows and exceeds some critical density, the
subsequent behaviors can be classified into three types.
The first one is the behavior shown in Fig.~\ref{f:main}, where the
hexagonal order and disorder alternately emerge.
The second behavior is entrainment into the large density for a large pump
intensity $F_0$.
When the local peak density reaches the uppermost branch of
Eq.~(\ref{ss}), the whole system is entrained into the uppermost branch
and the dynamics settle down there.
The third behavior occurs when the width $d$ of the pump beam is large,
which is shown in Fig.~\ref{f:local}.
After the hexagonal pattern is formed and destroyed at $t \simeq 400$ ps
in a manner similar to Figs.~\ref{f:main}(a)--\ref{f:main}(f), the
collapse and restoration of the peaks occur locally and sporadically,
which results in the pattern shown in
Figs.~\ref{f:local}(a)--\ref{f:local}(c).
This is in contrast with the dynamics in Fig.~\ref{f:main}, where the
order--disorder oscillation occurs synchronously in the whole system.
Figure~\ref{f:local}(d) shows a spatiotemporal image of the density
$|\psi_{\rm C}|^2$ on the dashed line in Fig.~\ref{f:local}(a).
Each density peak blinks periodically with a period $\simeq 4$ ps.
This may be related with the blinking phenomenon observed in
experiments~\cite{Baas}.
Since the peaks grow and collapse at various places, the norms $N_{\rm C}$
and $N_{\rm X}$ fluctuate around their averages, as shown in
Fig.~\ref{f:local}(e).
There is also a case in which several cycles of the synchronous
order--disorder oscillations shown in Fig.~\ref{f:main} occur, followed by
the local oscillations shown in Fig.~\ref{f:local}.

\begin{figure}[tbp]
\includegraphics[width=8cm]{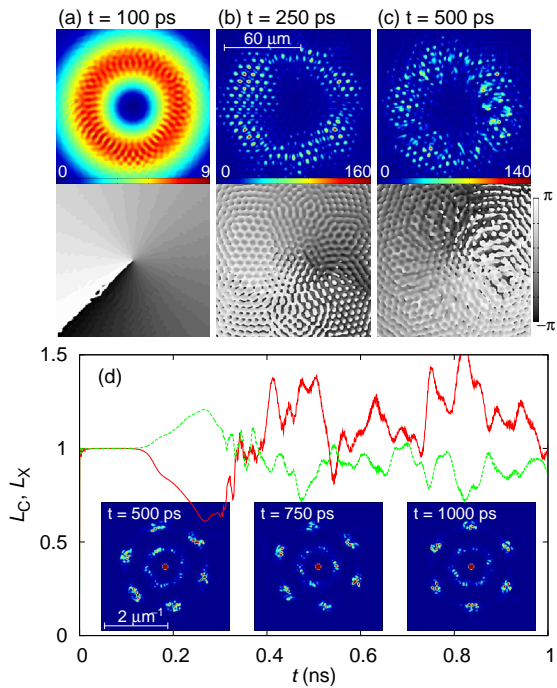}
\caption{
(color online) (a)--(c) Snapshots of the density $|\psi_{\rm C}|^2$ and
phase ${\rm arg} \psi_{\rm C}$ profiles for Laguerre--Gaussian pumping.
The pump parameters in Eq.~(\ref{LG}) are $\hbar F_0 = 21$ meV and $d =
50$ $\mu{\rm m}$.
The field of view of each panel is $120 \times 120$ $\mu{\rm m}$.
(d) Time evolution of the normalized angular momenta $L_j = -i \int
\psi_j^* \partial_\theta \psi_j d\bm{r} / \int |\psi_j|^2 d\bm{r}$ of 
photons ($j = C$, red solid line) and excitons ($j = X$, green dashed
line).
The insets show Fourier-transformed profiles, where the field of view
is $4 \times 4$ $\mu{\rm m}^{-1}$.
Except for the pump function, the parameters are the same as those in
Fig.~\ref{f:main}.
See supplemental material~\cite{movie}.
}
\label{f:vortex}
\end{figure}
We next examine polariton pumping with an angular momentum using a
Laguerre--Gaussian beam.
The pump function is given by
\begin{equation} \label{LG}
F(\bm{r}) = F_0 \frac{x + i y}{d} \exp\left[ -(x^2 + y^2) / d^2 \right].
\end{equation}
The pumped polaritons form a ring-shaped density profile with a phase
$\theta = {\rm arg} (x + i y)$ and the modulational instability
grows along the ring, as shown in Fig.~\ref{f:vortex}(a).
The triangular lattice initially has some mismatches along the ring, as
shown in Fig~\ref{f:vortex}(b), since the lattice grows from different
points on the ring.
Subsequently, in a manner similar to Fig.~\ref{f:local}, the peaks
collapse and revive locally (Fig.~\ref{f:vortex}(c)).
The angular momenta and the Fourier-transformed profiles shown in
Fig.~\ref{f:vortex}(d) exhibit an interesting behavior.
Although the angular momentum per particle is $\sim \hbar$ on average,
from which the angular velocity for a typical radius $R \simeq 40$
$\mu{\rm m}$ is estimated to be $\dot\theta \simeq \hbar / (m_{\rm LP}
R^2) \simeq 1.8$ ${\rm rad} / {\rm ns}$, the hexagonal pattern in
Fig.~\ref{f:vortex}(d) does not rotate; rather, it drifts slightly in the
opposite direction (clockwise).
This indicates that the hexagonal crystalline order is almost irrotational
despite the rotation of the superflow~\cite{Saito}; this may be regarded
as a supersolid-like behavior.

In experiments, a disorder potential in a sample typically has an energy
of at least $\sim 0.1$ meV and a wavelength of $\sim 10$ $\mu{\rm
m}$~\cite{Roumpos}.
We have numerically confirmed that dynamics similar to
Figs.~\ref{f:main}--\ref{f:vortex} are observed even when the disorder
potential is added to the rhs of Eq.~(\ref{GPc}).
Under continuous pumping of polaritons such as in the present situation,
it is experimentally difficult to perform time-resolved measurements.
Time-integrated imaging will reduce the hexagonal order, since its
direction is variable (see Figs.~\ref{f:main}(e) and \ref{f:main}(i))
because of the rotational symmetry of the Hamiltonian.
However, in reality, the disorder potential breaks the rotational symmetry
and fixes the direction of the hexagonal order, which enables us to use
time-integrated imaging to identify the hexagonal order.
The order--disorder oscillation shown in Fig.~\ref{f:main} may also be
detected by taking spatial and temporal correlations~\cite{Amo11}.

\section{Conclusions}
\label{s:conc}

We have investigated the dynamics of an exciton--polariton
superfluid coherently pumped in a semiconductor microcavity.
Under appropriate pumping, a hexagonal crystalline order is formed,
destroyed, and restored; this cycle is repeated many times
(Fig.~\ref{f:main}).
Such a spontaneous oscillation between ordered and disordered states is a
unique phenomenon in a simple nonequilibrium open system.
We have also shown that superflow can coexist with crystalline order at
rest (Fig.~\ref{f:vortex}), which is reminiscent of supersolidity.
The continuously pumped exciton--polariton system thus affords a quantum
dissipative system that is promising for studying novel dissipative
structures.

\begin{acknowledgements}
This work was supported by Grants-in-Aid for Scientific
Research (No.\ 22340116 and No.\ 23540464) from the Ministry of Education,
Culture, Sports, Science and Technology of Japan.
\end{acknowledgements}

\end{document}